\documentclass[journal=jacsat,manuscript=article]{achemso}

\usepackage[version=3]{mhchem} % Formula subscripts using \ce{}
\usepackage{multirow}

\author{M. Titze}
\affiliation[SNL]{Sandia National Laboratories, Albuquerque, NM, 87123, USA}
\email{mictitz@sandia.gov} %% email address is required
\author{H. Byeon}
\affiliation[CINT]{Center for Integrated Nanotechnologies, Sandia National Laboratories, Albuquerque, NM, 87123, USA}
\author{A. R. Flores}
\affiliation[SNL]{Sandia National Laboratories, Albuquerque, NM, 87123, USA}
\author{J. Henshaw}
\affiliation[CINT]{Center for Integrated Nanotechnologies, Sandia National Laboratories, Albuquerque, NM, 87123, USA}
\author{C. T. Harris}
\affiliation[CINT]{Center for Integrated Nanotechnologies, Sandia National Laboratories, Albuquerque, NM, 87123, USA}
\author{A. M. Mounce}
\affiliation[CINT]
{Center for Integrated Nanotechnologies, Sandia National Laboratories, Albuquerque, NM, 87123, USA}
\author{E. S. Bielejec}
\affiliation[SNL]{Sandia National Laboratories, Albuquerque, NM, 87123, USA}

\title{Towards Deterministic Creation of Single Photon Sources in Diamond using In-Situ Ion Counting}

\begin{document}
%TC:ignore
\begin{abstract}
%Defect-based single photon emitters in diamond have sparked interest as a potential platform for scalable quantum information applications. Among other challenges, the low activation yield of impurity ions to optically active defect centers prevent widespread use of diamond as a host material for quantum applications. While various techniques have been developed to increase the yield of implanted ions to optically active defects, such as C coirradiation, e-beam irradiation and localized laser heating, no one solution can deterministically create single photon emitters. Even with an increased yield, high confidence of the number of implanted ions is required to overcome the Poisson statistics for low number of ions.
We present an in-situ counted ion implantation experiment reducing the error on the ion number to 5 \% enabling the fabrication of high-yield single photon emitter devices in wide bandgap semiconductors for quantum applications. Typical focused ion beam implantation relies on knowing the beam current and setting a pulse length of the ion pulse to define the number of ions implanted at each location, referred to as timed implantation in this paper. This process is dominated by Poisson statistics resulting in large errors for low number of implanted ions. Instead, we use in-situ detection to measure the number of ions arriving at the substrate resulting in a two-fold reduction in the error on the number of implanted ions used to generate a single optically active silicon vacancy (SiV) defect in diamond compared to timed implantation. Additionally, through post-implantation analysis, we can further reduce the error resulting in a seven-fold improvement compared to timed implantation, allowing us to better estimate the conversion yield of implanted Si to SiV. We detect SiV emitters by photoluminescence spectroscopy, determine the number of emitters per location and calculate the yield to be $2.98 + 0.21 / - 0.24 \%$. Candidates for single photon emitters are investigated further by Hanbury-Brown-Twiss interferometry confirming that $82 \%$ of the locations exhibit single photon emission statistics. This counted ion implantation technique paves the way towards deterministic creation of SiV when ion counting is used in combination with methods that improve the activation yield of SiV.
\end{abstract}
%TC:endignore

%%%%%%%%%%%%%%%%%%%%%%%%%%  body  %%%%%%%%%%%%%%%%%%%%%%%%%%
\section{Introduction}
Solid-state single photon emitters (SPE) have generated significant interest for their potential use in quantum applications \cite{Aharonovich2016}. A variety of host systems have been investigated for SPEs, such as silicon-on-insulator, two-dimensional materials and wide-bandgap materials such as SiC, AlN, GaN or diamond \cite{He2015,Chakraborty2015,Koperski2015,Srivastava2015,Peyskens2019,Hollenbach2020}. An advantage of SPEs hosted in wide-bandgap semiconductors is the potential for (near) room temperature operation. Color centers in diamond are very promising candidates, since SPEs with various optical properties can be hosted in one material system \cite{DitalaTchernij2018,Aharonovich2011,Bradac2019}. For example the nitrogen-vacancy has millisecond spin-coherence lifetime while group-IV color centers benefit from high emission rates \cite{Trusheim2019}. The silicon-vacancy (SiV) benefits from a large Debye-Waller factor at room-temperature \cite{Neu2013} whereas heavier group-IV color centers such as the germanium-vacancy (GeV), tin-vacancy (SnV) and lead-vacancy (PbV) promise high emission rate in combination with long spin-coherence lifetime due to larger ground state splitting \cite{Schroder2017,Trusheim2019,Trusheim2020,DeSantis2021}. Fabrication of these group-IV defects is often performed by either broad beam or Focused Ion Beam (FIB) implantation, where the number of implanted ions is controlled by measuring the beam current prior to implantation, followed by high temperature activation annealing \cite{Sipahigil2016,Wan2020}. For our purpose we will refer to these passive techniques as timed implantation.

%One issue in creating devices with color centers in diamond is the low activation yield, typically between $2-5 \%$ for group-IV defects \cite{Wan2020}. For Si, the low activation yield is ascribed to the lack of suitable vacancies in the local environment of the Si ions preventing the creation of a Si divacancy complex \cite{Schroder2017,Lagomarsino2018}. In addition to low activation yield
Creation of a single SiV requires implantation of a small number of ions leading to significant uncertainty in the number of ions implanted due to Poisson statistics inherent to ion emission \cite{Singh2016,Abraham2016,Pacheco2017}. To create a single SiV without advanced activation techniques \cite{Schroder2017}, implantation of on average 30 ions is required. For timed implantation this leads to an error of $\pm 5.5$ ions which is $\pm 18 \%$ of the total ions implanted. With improvements in the activation yield of SiV, the relative uncertainty in the number of ions will be higher since for implantation of $N$ ions the error on the ion number is $\sqrt{N}$ resulting in the need for in-situ counting.

We previously developed a counting system \cite{Abraham2016,Pacheco2017} which enables detection of ions as they are implanted into the diamond using in-situ charge measurements based on the detection of generated electron-hole (e-h) pairs. In this paper, we have extended this technique resulting in a counted array of SiV centers where we have reduced the implantation error by a factor two and with post implantation analysis we can further reduce the error to a total factor of seven as compared to timed implantation. Photoluminescence (PL) measurements confirm on average one emitter per location and show Poisson statistics of SiV activation. This approach allows the error of the SiV activation yield to be improved to $2.98 + 0.21 / - 0.24 \%$, a three-fold improvement of the uncertainty of the activation yield \cite{Schroder2017,Lagomarsino2018,Hunold2021}. Hanbury-Brown-Twiss (HBT) measurements were performed on a subset of SPEs with $82 \%$ showing single photon emission statistics. The remaining emitters showed multiphoton but still non-classical emission statistics, likely from NVs formed from native N overlapping with the SiV emitter location.

\section{Results}
\subsection{Counted Ion Implantation}
\begin{figure}
    \centering
    \includegraphics[width=\textwidth]{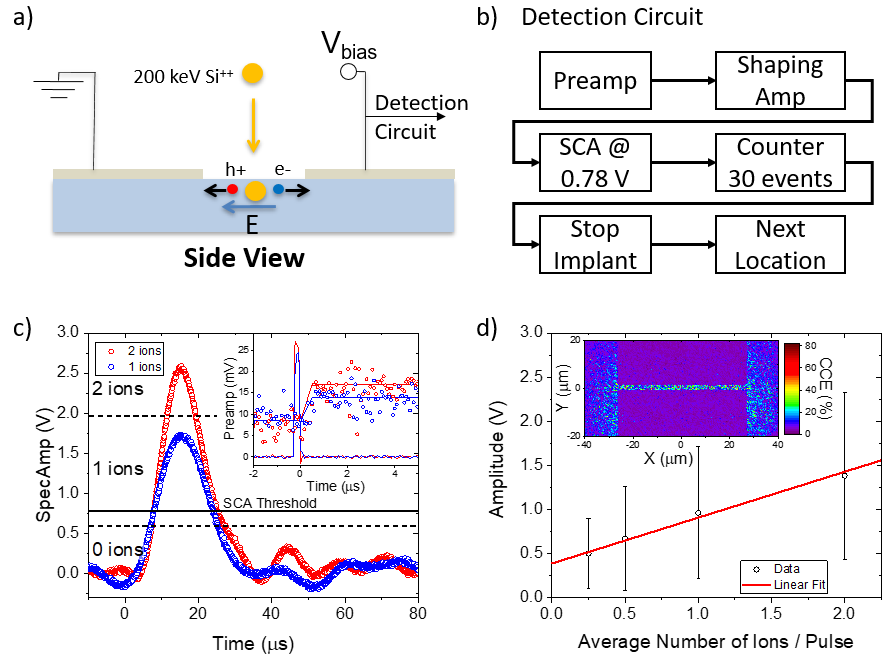}
    \caption{(a) Schematic diagram of the ion counting setup. (b) Detection flow diagram for deterministic implantation of a preset number of ions. (c) Main: Spectroscopy amplifier trace for pulses containing 1 and 2 ions. The solid black line is the SCA threshold at 0.78 V, the dashed lines show the bins for post-implantation analysis. Inset: Preamp signal corresponding to the spectroscopy amplifier traces shown in the main figure. The lines are a fit to the preamp signal. The ion beam is turned on during the time the trigger signal is high, denoted as the red and blue traces. (d) Main: Calibration curve taken from the average spectroscopy amplifier amplitudes of 10,000 ion pulses, showing the device response is linear for few ion implantation. Inset: IBIC map showing the active area of the detector and high CCE within the active area.}
    \label{fig:CountingSetup}
\end{figure}

We implant Si$^{++}$ ions with 200 keV landing energy using the A\&D nanoImplanter (nI) into 5 areas containing arrays of $4 \times 40$ ions where we counted in 10, 20 and 30 ions per location. Additional details on the ion implantation can be found in the Supplementary Information (SI). In-situ probes are used to measure the charge induced on probe pads deposited on the diamond. By biasing the probes we generate an electric field of 6 V/\textmu m, see Figure \ref{fig:CountingSetup}a. Figure \ref{fig:CountingSetup}b shows the signal path developed for in-situ counting. Each ion striking the diamond detector loses energy through both electronic and nuclear stopping generating an average of $1.2 \times 10^{4}$ e-h pairs per ion strike \cite{SRIMWebsite}. Details on the energy partitioning can be found in the SI. The sample bias separates the e-h pairs and induces charge in the probes which is then amplified through a charge-sensitive preamplifier outside the vacuum chamber. The preamplifier time trace for response to a single (blue) and two (red) ion strike is shown in the inset of Figure \ref{fig:CountingSetup}c, alongside fits to the preamplifier signal. The main graph in Figure \ref{fig:CountingSetup}c is the spectroscopy amplifier trace corresponding to the two shown preamplifier outputs. Also shown are the single channel analyzer (SCA) threshold as well as the bins used for post-implantation analysis, discussed later. Each e-h pair generates a signal of 0.64 \textmu V, which in combination with the number of e-h pairs per ion strike can be used to estimate the charge collection efficiency (CCE) of the detector. An ion beam induced charge collection (IBIC) map is shown as the inset in Figure \ref{fig:CountingSetup}d, this is collected by scanning the beam over a 80 \textmu m $\times$ 40 \textmu m area with an average of 2 ions/pulse. The CCE is determined to be ~83 \% and is nearly constant within the active region of the detector but rolls off rapidly outside the area of uniform electric field. %We attribute the relatively low CCE for a diamond detector to the small number of ions per pulse approaching the sensitivity limit of our detection scheme. 
The main graph in Figure \ref{fig:CountingSetup}d shows the calibration of our detector for few ions demonstrating that the detector response is linear for low ion doses. The error bar denotes one standard deviation of the measured values based on the average of 10,000 pulses.

To perform a deterministic implant, the implantation process must stop upon reaching a preset number of ions. We use a SCA with the upper threshold disabled connected to a counter to count ions at the sample. The lower-level threshold of the SCA is set to reject events that create an amplitude lower than 0.78 V (see Figure \ref{fig:CountingSetup}c) on the spectroscopy amplifier. The low threshold was set such that no background counts were detected with the beam off. Since the SCA cannot distinguish between implantation of one or more ions, we set the ion pulse length to $<0.1>$ ions/pulse, greatly reducing the number of pulses with more than 1 ion. For $<0.1>$ ions/pulse, 90.55\% of all pulses will contain zero ions, $9 \%$ of pulses will contain one ion and $0.45 \%$ of pulses are expected to contain more than one ion.

\begin{figure}
    \centering
    \includegraphics[width=\textwidth]{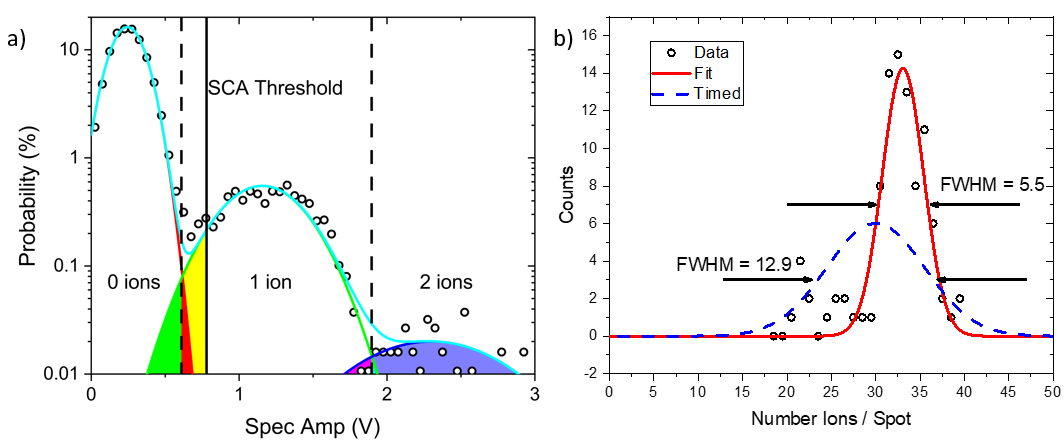}
    \caption{(a) Histogram of spectroscopy amplifier amplitude for all implantation events into one array. Gaussian fits to 0, 1 and 2 ion implantation events are shown in red, green, blue respectively with the cumulative fit shown as the cyan curve. The solid black vertical line denotes the SCA threshold used in in-situ ion counting and the dashed vertical lines show the bins for 0, 1 and 2 ion implantation events in our ex-situ analysis. (b) Histogram of ions implanted per location as determined from the individual spectroscopy amplifier traces. The red curve is a fit to the data from which we extract the FWHM. The dashed blue curve is a simulation assuming an average of 30 ions/spot with the Poisson limited standard deviation of 5.5 ions leading to a much broader distribution.}
    \label{fig:CountingResult}
\end{figure}

Figure \ref{fig:CountingResult}a shows a histogram of the detected spectroscopy amplifier amplitudes from the counting experiment, exhibiting three peaks corresponding to implantation of 0, 1 and 2 ions. We fit the peaks with Gaussian functions such that the area under the curve corresponds to the probability of implanting the respective number of ions, allowing us to extract error rates for false positives, false negatives and multiple ion implantation. To do so, we compare the area of each of the errors to the area of the peak corresponding to implantation of one ion in Figure \ref{fig:CountingResult}a. For false positives, we measure the overlap of the peak corresponding to implantation of zero ions to the right of the SCA threshold denoted as the solid black line and find less than 1 ppb false positives. To estimate false negatives, we use the yellow shaded area, which corresponds to an $8.6 \%$ likelihood of a single ion implantation event being incorrectly identified as a zero. From the area ratio of the peaks corresponding to implantation of zero and one ions we find the actual beam current during the experiment was $<0.112>$ ions/pulse, likely due to a slow beam drift. This more accurate measurement is used to calculate the error due to implantation of multiple ions. Since multiple ion implantation is not captured by the SCA. We expect an additional error of $5.8 \%$ of all pulses containing at least one ion to contain two or more ions, denoted as the magenta shaded area.
Adding these error rates together yields an overall error of $+ 14.4 / - 0\%$, a two-fold improvement over timed implantation.  The positive error rate results in over implantation, dominated by a combination of the false negative and the multiple implantation errors, and a negative error leads to fewer implanted ions. 

To further reduce the error in the implanted number of ions, we perform post-implantation analysis to measure the number of ions implanted per location. The dashed vertical lines in Figure \ref{fig:CountingResult}a denote the bins for post implantation classification of events into 0, 1 and 2 ions. After implantation the threshold for classification of an ion implantation event was changed to 0.6 V, from 0.78 V during the in-situ measurement, minimizing the overall error by finding a compromise between false negatives and false positives instead of deliberately rejecting all false positives. In Figure 2a, the false negatives $(2.3 \%)$ are denoted as the green shaded area and the false positives $(0.86 \%)$ are the red shaded area. Additionally, we added a second threshold at 1.96 V to capture implantation of multiple ions. The new error from implantation of multiple ions is denoted as the blue shaded area $(1.7 \%)$. Furthermore, some single ion implantation events will be classified as multiple ion events using the chosen binning, which leads to erroneously counting single ion events as multiple ion implantation, denoted as the cyan shaded area $(0.2 \%)$.
Adding these error rates yields an overall error of $+ 4.0 / - 1.1 \%$, a seven-fold improvement over timed implantation.

In Table \ref{tab:Errors} we compare the errors between timed, in-situ counting and post-implantation analysis for implantation of 30 ions. The total errors are the sum of all errors.

\begin{table}
 \centering
 \begin{tabular}[t]{c|c c c}
  Implant 30 ions & Timed & In-Situ & Post-analysis \cr
  \hline
  False Negative & - & 8.6 \% & 2.3 \% \cr
  False Positive & - & $< -1$ ppb & -0.86 \% \cr
  Multiples & - & 5.8 \% & 1.7 \% \cr
  Single as Double & - & - & -0.2 \% \cr
  \hline
  Total & $+18.3 / -18.3$ \% & $+14.4 / -0$ \% & $+ 4.0 / - 1.1$ \%
 \end{tabular}
 \caption{Overview of the error on the number of implanted ions for different techniques. For timed implantation the reported error is one standard deviation.}
 \label{tab:Errors}
\end{table}

In the experiment, we set the counter to stop implantation after 30 ions/spot. Through a combination of the in-situ measurement with post-implantation analysis we can now determine the number of ions implanted into each location. In Figure \ref{fig:CountingResult}b we show the histogram of implanted ions per spot of the ensemble of all implanted spots within one array. The points are obtained from post-implantation analysis of the in-situ data and fit well to a Gaussian, the red curve. From this we determine the average number of implanted ions per spot to be 33.8, $13 \%$ higher than the preset of 30 and at the high end of what is expected from the in-situ counting error of $+14.4 \%$. We expected this over implantation as the all-positive errors from in-situ counting (see Table \ref{tab:Errors}) will not be caught by the detection electronics. The blue dashed curve shows the distribution of ions in a timed experiment assuming an average of 30 ions/spot with a standard deviation of $18 \%$ showing that the counting experiment reduces the uncertainty in the number of implanted ions.

\subsection{Optical Measurement}
\begin{figure}
     \centering
     \includegraphics[width=\textwidth]{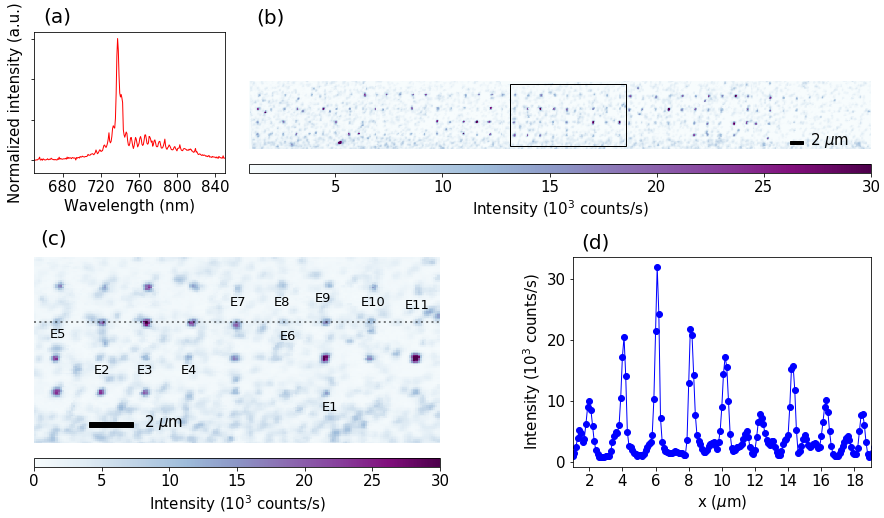}
     \caption{(a) PL Spectrum of a single SiV at room temperature. (b) PL map of an in-situ counting array of SiV. (c) Zoomed in view of black box in (b) showing individual emitters with intensity fluctuations. The emitters labeled E1-E11 are selected as SPE candidates. (d) Linecut along the dashed black line in (c) showing a pitch of 2 \textmu m between implant locations and the PL intensity variation.}
     \label{fig:Array}
\end{figure}

\begin{figure}
    \centering
    \includegraphics[width=\textwidth]{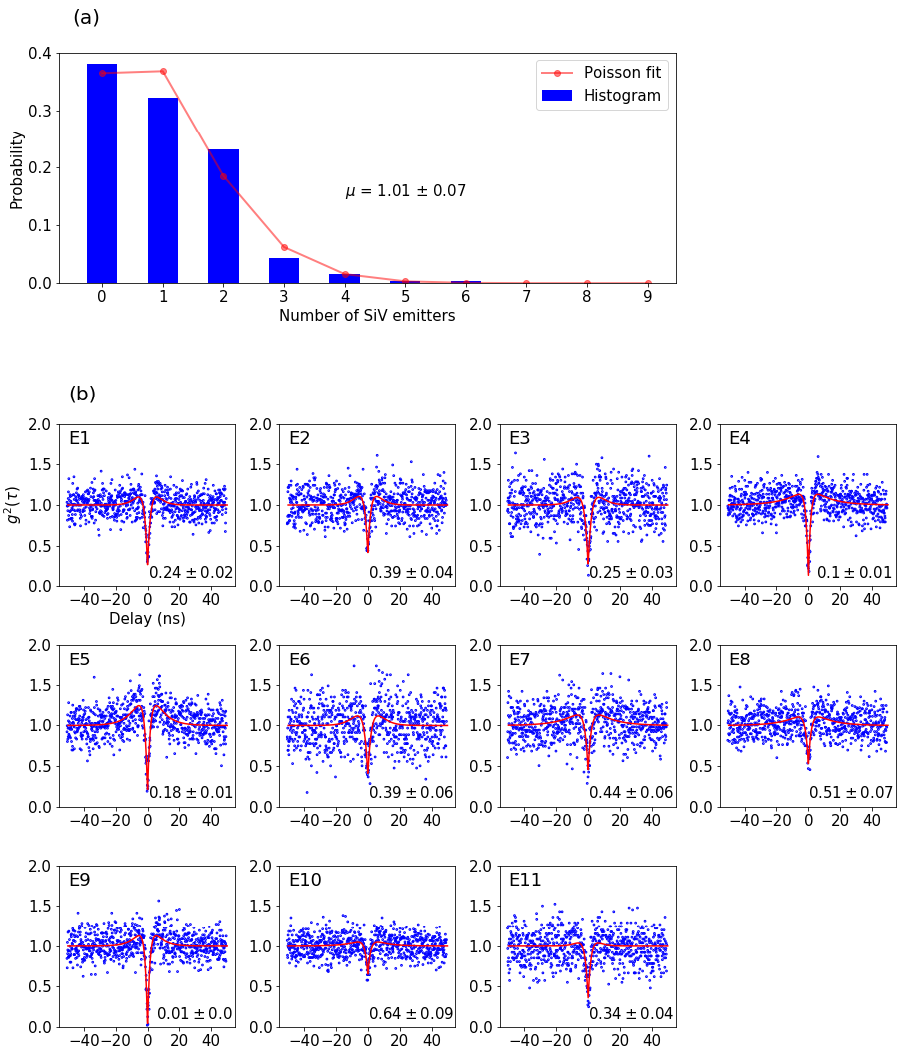}
    \caption{(a) Histogram showing the probability of finding a certain number of emitters at each location with a fit assuming a Poisson distribution. A total of 320 implant locations are used for the histogram. The expectation value of 1.01 SiV / site is extracted from the fit. (b) Second-order correlation measurement of the emitters labeled in Figure \ref{fig:Array}c. The blue points are the measured data and the red line is a fit using Equation \ref{eq:g2}.}
    \label{fig:g2}
\end{figure}

We collected emission spectra using a spectrometer to verify the detected PL is due to SiV emission and perform HBT interferometry to confirm SPE statistics. Details on the setup are in the SI. The SiV spectrum in Figure \ref{fig:Array}a exhibits the characteristic zero-phonon line (ZPL) emission at 738 nm \cite{Aharonovich2011}. Since we implanted on average 33.8 ions/spot, we can calculate the activation rate of the implanted Si to SiVs by counting the number of SiVs created. The implantation was performed with a pitch of 2 \textmu m in both horizontal and vertical direction. The resulting array is shown in Figure \ref{fig:Array}b, with a zoomed in view of the area highlighted by the black box in Figure \ref{fig:Array}c and a linecut along the dashed line in Figure \ref{fig:Array}d. To calculate the SiV yield we define a grid over the array and at each point select a circular area with 2 \textmu m diameter. We fit each area to a two-dimensional Gaussian to extract the amplitude of the PL. As the count rate for one SiV in the confocal microscope was found to be $10.7 \pm 0.2$ kcps, we use the bins $0 \pm 5, 10 \pm 5, 20 \pm 5$, etc. kcps to classify locations into containing 0, 1, 2, etc. SiV. Figure \ref{fig:g2}a shows the histogram acquired by this binning corresponding to the respective probability of finding a certain number of emitters at any location in the array. The histogram data was fitted to a Poisson distribution and we find that the average number of emitters per location is $1.01 \pm 0.07$ SiV, which translates to a conversion yield of $2.98 + 0.21 / - 0.24 \%$ from implanted Si to SiV.

In addition to the PL measurement we measure the second-order autocorrelation $g^{(2)}(t)$ using HBT on the spots with less than 25 kcps in Figure \ref{fig:Array}b. To obtain an accurate measurement of $g^{(2)}(t)$, we perform background correction \cite{Brouri2000} as
\begin{equation*}
    g^{(2)}(t) = \frac{g_{exp}^{(2)}(t) - 1 + \rho^2}{\rho^2}.
\end{equation*}
Here $g_{exp}^{(2)}(t)$ is the experimental measurement of the autocorrelation function and $\rho = \frac{E-B}{E}$ is the ratio of the single photon count rate to the total count rate, where $E$ and $B$ denote the count rate at the emitter and background locations, respectively.

Figure \ref{fig:g2}b shows the normalized autocorrelation histogram of emitters E1 - E11 with background correction. All spots exhibit non-classical characteristics of the fluorescence, meaning $g^{(2)}(t=0) < 1$. We observe a clear bunching effect in the vicinity of $t = 0$ from most of the spots, which indicates an additional shelving state. Thus, the data is fit with a three-energy level model \cite{Fuchs2015}
\begin{equation} \label{eq:g2}
    g^{(2)}(t) = 1 - \frac{1}{N} + [1 - (1+a) e^{-|t|/t_1} + a e^{-|t|/ t_2}]/N ,
\end{equation}
where N is the number of SPE, $a$ is the ratio of emission from the shelving to the excited state, $t_1$ is the excited state lifetime and $t_2$ the shelving state lifetime. In the ensemble shown, we measure $t_1 = 2.33 \pm 0.7$ ns and $t_2 = 6.23 \pm 3.55$ ns, close to values reported in the literature \cite{Wang2005}. We observe that $82 \%$ of the emitters identified as SPE based on PL countrate exhibit SPE statistics indicated by $g^{(2)}(t=0) < 0.5$. For the remaining $18 \%$ of the spots (for example E8 and E10), the measured values of $g^{(2)}(0)$ are between 0.5 and 0.7, indicating two emitters in the same confocal field of view and spectral region.

\section{Discussion}
%We illustrate a technique for high confidence ion implantation by using high throughput ion counting. Through a combination of in-situ ion detection and ex-situ analysis of the implanted ions we reduce the uncertainty in the implant at each location by more than seven-fold when compared to a timed experiment.

Overall, the error in the implanted number of ions is due to the decribed errors (false positive, false negative, multiple ions, miscounting single as multiple) in our counting with additional errors due to a change in the number of ions/pulse, ascribed to beam drift. While we set the pulse length to $<0.1>$ ions/pulse, from the analysis of the area ratio between the peaks corresponding to implantation of 0 and 1 ions in Figure \ref{fig:CountingResult}a, we find the implantation rate of $<0.112>$ ions/pulse. In a timed experiment, this error would have remained undetected as we typically cannot measure beam current during implantation.

PL maps show that we create an incomplete array of SPE using in-situ ion counting, which is expected due to Poisson statistics of SiV formation. HBT measurements are performed on emitters classified as SPE by their PL countrate to confirm they are SPEs and we find that $18 \%$ of the spots contain two emitters. We note that in our experiments SiV are excited via a green 532 nm laser, an energy significantly higher than the energy required to excite SiV ZPL at 738 nm. Most notably, the NV ZPL is at 637 nm, such that we can expect our laser to excite native NV centers near the SiV locations. In particular, emission from NVs is most likely to affect $g^{(2)}(0)$ of SiV since the NV emission phonon sideband spectrum overlaps with SiV \cite{Simpson2009, Babinec2010, Aharonovich2010}. To determine the average NV separation we need to consider both the N density in an electronic grade diamond substrate, $0.1 - 1$ ppb, and the expected conversion yield from N to NV. The conversion yield has been previously experimentally measured to be $0.03 - 45 \%$ as a function of N energy \cite{Pezzagna2010}. For the range of Si ion energies implanted here, we can have NV conversion yield between $10 - 45 \%$, limited by the depth of created vacancies or the total number of vacancies available, respectively. We summarize the expected NV separation for the different assumptions of native N content and conversion yield in Table \ref{tab:Separation}.

\begin{table}[]
 \centering
  \begin{tabular}{c c|c c}
    \multicolumn{2}{c}{\multirow{2}{*}{\shortstack{Average NV \\ Separation (nm)}}} & \multicolumn{2}{c}{Native N Content} \cr
    & & 1 ppb & 0.1 ppb \cr
    \hline
   \multirow{2}{*}{\shortstack{Conversion \\ Yield}} & 45 \% & 230 & 497 \cr
    & 10 \% & 381 & 820 \cr
  \end{tabular}
 \caption{Average NV spacing for different assumptions of native N content and N to NV conversion yield.}
 \label{tab:Separation}
\end{table}

We observe two emitters in $18 \%$ of the locations classified as SPE based on PL countrate. Taking into account the laser excitation volume, we expect a mean separation between emitters of 340 nm. From the table above, this suggests a native N content of 1 ppb with an activation yield slightly above $10 \%$.

%Assuming a mean NV separation of 340 nm as expected from 1 ppb N concentration and a NV conversion yield just above $10 \%$, the probability of finding an NV within the detection volume is $18 \%$, matching the measured $18 \%$ of locations that contain more than one emitter.

Conversely, we could indeed have a second low-brightness SiV emitter at the locations where we detect $g^{(2)}(0) > 0.5$. The conversion yield then would be $3.52 \%$ instead of the $2.98 \%$ yield we measure based on PL measurements. In this regard, a lower energy excitation laser which cannot excite NVs \cite{Wang2005} or low-temperature measurements to suppress the NV phonon sideband would be more suitable for SiV $g^{(2)}(t)$ measurements.

\section{Conclusion}
We created arrays of SiV emitters in diamond using an in-situ ion counting technique. By using a post-implantation analysis of the acquired counting data, we can further reduce the error bar on the number $N$ of implanted ions from $\sqrt{N}$ ions to only $+4 / -1.1 \%$ of the total number of implanted ions, a seven-fold improvement over timed implantation. We measure the yield of optically active SiV from the number of implanted Si to be $2.98 + 0.21 / - 0.24 \%$ from PL brightness. However, $18 \%$ of emitter locations identified as single emitters based on the PL brightness do not exhibit SPE statistics as measured by HBT interferometry. We tentatively ascribe this to NV emission. The low activation yield of implanted Si ions to optically active SiV prevents us from creating a fully deterministic array of single SiV. In the future, combining our in-situ ion counting technique with laser or electron irradiation, the yield may be improved from $\approx3 \%$ to $\approx25 \%$, making in-situ ion counting a crucial component for the creation of deterministic single SiV emitters \cite{Schroder2017}. Additionally, a different implantation ion or different defect center which generates emitters with much higher yield in diamond, or other substrates may pave the way for fully deterministic placement of SPEs.

\section{Acknowledgements}
This work was performed, in part, at the Center for Integrated Nanotechnologies, an Office of Science User Facility operated for the U.S. Department of Energy (DOE) Office of Science. Sandia National Laboratories is a multimission laboratory managed and operated by National Technology \& Engineering Solutions of Sandia, LLC, a wholly owned subsidiary of Honeywell International, Inc., for the U.S. DOE’s National Nuclear Security Administration under contract DE-NA-0003525. The views expressed in the article do not necessarily represent the views of the U.S. DOE or the United States Government.

%\section{References}
\bibliography{DiamondCounting}
%\nocite{*}
%TC:ignore
\pagebreak
\begin{center}
 \section{Supplementary Information for Towards Deterministic Creation of Single Photon Sources in Diamond using In-Situ Ion Counting}
 M. Titze, H. Byeon, A. R. Flores, J. Henshaw, C. T. Harris, A. M. Mounce and E. S. Bielejec
\end{center}

\subsection{Sample Preparation}
We used electronic grade diamond from Element 6 with approximately 1 ppb N for these experiments. The as-received diamond samples were cleaned in acetone and isopropyl alcohol and then subjected to a piranha 3:1 H$_2$SO$_4$:H$_2$O$_2$ bath for 10 minutes to remove organic residue. To remove surface polishing defects, we used a CF$_4$ reactive ion etch to remove the top $5-10$ \textmu m of the diamond. We then defined alignment markers using an Al hard mask. After a metal lift-off, a fluorine etch defined the alignment markers, with the remaining Al stripped via an Al wet etchant. Probe pads were defined using a subsequent Al metallization. After implantation, samples were annealed at $1 \times 10^{-8}$ Torr. The heat treatment consisted of ramping and staying at 400\textdegree C for 4h, then ramping to 800\textdegree C for another 4h and finally ramping to 1200\textdegree C for 2h. The heating ramp rate was kept constant at 100\textdegree C/h for all ramp and cooldown steps. After annealing, the samples were subjected to a tri-acid clean using a 1:1:1 ratio of H$_2$SO$_4$:HNO$_3$:HClO$_4$ in a boiling flask and refluxed with a water-cooled condenser for 1.5h at $250$\textdegree C. A final piranha clean was performed to remove residual carbon and to oxygen-terminate the diamond surface.
\subsection{Ion Implantation}
Ion implantation was performed using an A\&D nanoImplanter (nI), a 100 kV focused ion beam system enabling high-resolution patterning and single ion implantation. In-situ Kleindiek electrical probes were used to make electrical conacts \cite{Abraham2016,Pacheco2017}. We utilized liquid metal alloy ion source (LMAIS) made of Au$_{77}$Ge$_{14}$Si$_{9}$ with a Wien filter to pick out 28-Si$^{++}$ for implantation. The double-charged Si ions have a landing energy of 200 keV at 100 kV acceleration potential with a range of 130 $\pm$ 21 nm in diamond according to SRIM simulations \cite{SRIMWebsite}. The ions lose energy and create e-h pairs in the process. The total energy loss contributions are shown in Supplementary Figure \ref{fig:SRIM} and are summarized as: (1) Direct ionization generating $6.8 \times 10^{3}$ e-h pairs, (2) Ions recoiling lattice atoms generating an additional $4.9 \times 10^{3}$ e-h pairs, and (3) the remainder of the beam energy is lost through phonons not leading to formation of an e-h pair. The ion beam was pulsed for 225 ns, corresponding to an average of 0.1 ions per pulse based on a measured beam current using the Faraday cup on the sample holder. Probe pads were used to bias the diamond and electron-hole (e-h) pairs generated by the ion beam induced a charge on the probe pads which was then amplified by a charge sensitive preamplifier (Amptek A250CF). The signal was then routed to a spectroscopy amplifier (Ortec 672) and analyzed using an oscilloscope (Tektronix MSO58) as well as a single channel analyzer (Ortec 551).

\begin{figure}
    \centering
    \includegraphics[width=\textwidth]{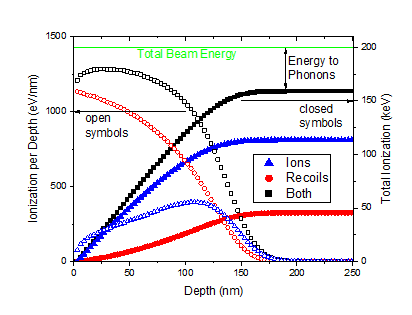}
    \caption{Energy loss to electronic stopping as a function of depth into the diamond substrate. The closed (open) symbols show the differential (integral) energy loss. The triangles denote energy lost due to direct ionization while the circles denote energy lost from recoiling atoms. The squares show the total energy loss to ionization measured in the counting experiment. The total beam energy is denoted as the solid green line and the difference between total beam energy and the total ionization is lost to phonons which cannot be detected by our setup.}
    \label{fig:SRIM}
\end{figure}

\subsection{Optical Measurements}
2D micro-photoluminescence (PL) scans of SiVs in diamond were carried out in a home built confocal microscope using continuous-wave 532 nm green laser (Sprout, Lighthouse Photonics) at 4 mW. The laser underwent a reflection at a 550 nm dichroic mirror and was then focused with an oil immersion objective (M = 60$\times$, NA = 1.42, Olympus). High resolution scanning at the sample was achieved using a XYZ piezo stage (NanoMax300, Thorlabs). PL from the SiV emitters was filtered through the 550 nm dichroic mirror, focused through a 50 \textmu m pinhole, filtered through a 740 nm single-band pass filter (FF01-740/13-25, Semrock) and then guided to a free-space Hanbury-Brown-Twiss (HBT) interferometer used for photon correlation measurements. The HBT setup is composed of two free-space single photon counting modules (SPCM-AQRH-14, Excelitas) and a time-correlated photon counting module (PicoHarp 300). During the HBT measurement time correlation data was acquired within one minute and repeated 100 times per emitter location. Between the one-minute acquisitions the microscope was refocused to optimize photon counts with the x,y and z positioners using the Nelder-Mead simplex method if the photon count rate dropped below $90 \%$ of the initial value.
%TC:endignore

SAND \#: SAND2021-15146 O
\end{document}